\documentclass{article}

\usepackage[latin1]{inputenc}

\usepackage{latexsym}
\usepackage{graphicx}

\newtheorem{axiom}{Axiom}
\newtheorem{theorem}{Theorem}
\newtheorem{defin}{Definition}
\newtheorem{remark}{Remark}

\title{Exact Philosophy of Space-Time
  \thanks{Facultad de Ciencias Astronómicas y Geofísicas ---
    Universidad Nacional de La Plata, ARGENTINA.}\\
       \Large\emph{In honour of Mario Castagnino}}
\author{H\'ector Vucetich}

\begin{document}
\maketitle
\bibliographystyle{unsrt}

\begin{abstract}
  Starting from Bunge's (1977) scientific ontology, we  expose a
  materialistic relational theory of space-time, that carries out the
  program initiated by Leibniz, and provides a protophysical basis
  consistent with any rigorous formulation of General
  Relativity. Space-time is constructed from general concepts which
  are common to any consistent scientific theory and they are
  interpreted as emergent properties of the greatest assembly of
  things, namely, the world.
\end{abstract}

\section{Introduction}
\label{sec:Intro}

All disciplines in modern science take the notions of space and time
for granted: physics describes elementary particles as
objects with wave functions of space and time, chemistry deals with
flows of reactants, ecology studies the wandering of plankton in
the multitudinous sea and sociology describes the interactions of
neighboring cultures along their history. 

But for all science space and time are primitive concepts, even for
General Relativity that associates the metric structure of space-time
with the gravitational field. Indeed, the question ``What is
space-time?'' belongs to protophysics: the branch of ontology dealing
with the basic assumptions in physics.

The ontological status of space and time has been a  subject
of debate for physicists and philosophers during the last 400
years. The kernel of this debate has been the confrontation between
two antagonic positions: absolutism and relationalism. The former,
held by Newton in his famous discussion with Leibniz (mediated by
S. Clarke) \cite{Alexander} was stated by him in his \emph{Principia}
\cite{NewtonP} 
\begin{quotation}
  Absolute time, true and mathematical, in itself and by its own
  nature, flows evenly without relation to any external thing.

  Absolute space, by its own nature, without relation with any
  external thing, stays always identical and motionless.
\end{quotation}

Thus, for absolutists, space-time is the stage where the drama of
nature is enacted; i.e. the absolutist position considers space-time as
much a thing as planets or electrons are: a physical entity endowed
with concrete properties. A modern version of the absolutist position
has been held by by J. Wheeler in his geometrodynamical approach to
physics \cite{MThW}.

 The relationalist position instead asserts that space-time is not a
thing but a complex of relations among physical things. In Leibniz's
words \cite{Alexander}:
\begin{quote}
  I have said more than once that I hold space to be something merely
  relative, as time is; that I hold it to be an order of coexistents,
  as time is an order of successions. 
\end{quote}
In our theatrical analogy, relationalists consider space-time as a
pattern weaved by the actors.

An important consequence of Leibniz's ideas is that if space-time is
not an ontological primitive, then it should be possible to construct
it starting from a deeper ontological level. That is to say, the
spatiotemporal relations should be definable from more fundamental
relations. There have been several attempts to analyze the relational
nature of space-time, both subjective and phenomenological (e.g.
\cite{CarLAW,Basri}) and objective and realistic
\cite{BungeMayne,Bunge3}.

In this paper we paper I shall present a simplification and
streamlining of a relational theory of space-time \cite{esptim}, based
on the scientific and realistic ontology of Bunge
\cite{Bunge3,Bunge4}.%
\footnote{The main simplifications with respect to \cite{esptim} are
  the introduction of Axiom \ref{axm:UnivInt}, the use of the notion of
  simultaneity to analyze clocks and of the axiom of distances to build
  uniformities. Also, a general rearranging of the theory of space has
  shortened the overall presentation.} 

The choice of Bunge's approach, which only assumes hypothesis common
to all science, is because a deductive theory of space-time cannot be
built with blocks alien to the physical science (such as cognoscent
subjects or sensorial fields) in order to be compatible with
contemporary physical theories.

 The theory is presented in an axiomatic way although
we shall limit ourselves in this paper to an informal presentation%
\footnote{On the advantages of the axiomatic method see \cite{AxQM1}
and references therein.}.

The ontological theory of space-time is a nice example of the
interaction of science (mainly physics) and philosophy. Indeed, the
hypothesis used to build space-time will be suggested by scientific
observation, leading to a consolidation of its foundations.

\section{Ontological summary}
\label{sec:Ontology}

 In this section we give a brief synopsis of the ontological
presuppositions that we take for granted in our theory. For greater
detail see \cite{Bunge3,Bunge4,AxQM2}. The basic 
statements of the ontology can be formulated as follows:

\begin{enumerate}

\item There exist concrete objects $x$, named {\em  things.} The set of
all the things is denoted by $\Theta$.

\item Things can juxtapose ($\dot{+}$) and superimpose
($\dot{\times}$) to give new things. Juxtaposition as superimposition
satisfy a Boolean algebra structure.

\item The \emph{null thing} is a fiction equal to the superimposition
  of all things.
  \[ \Diamond = \dot{\prod_{x\in\Theta}} x\]

\item Two things are separated if they do not superimpose:
\[
x \wr y \Leftrightarrow x \dot{\times} y = \Diamond
\]
Non-separated things are called united.

\item Let $T$ a set of things. The {\em aggregation} of $T$ (denoted $[T]$)
is the supremum of $T$ with respect to the operation $\dot{+}$.

\item The world $(\Box)$ is the aggregation of all things: 
\[ 
\Box = [\Theta] \Leftrightarrow (x \sqsubset \Box \Leftrightarrow x \in
	\Theta) 
\] 
where the symbol `$\sqsubset$' means `to be part of'. 

\item All things are made out of basic things $x\in {\Xi}\subset
\Theta$  by means of juxtaposition or superimposition. The basic
things are elementary or primitive:
\[ 
(x,y \in {\Xi}) \wedge (x \sqsubset y) \Rightarrow x=y 
\]

\item Things $x$ have {\em properties} $P(x)$. These properties can be
intrinsic or relational.

\item A property $p \in P(x)$ of a thing $x$ is called {\em hereditary} if
some of the components of $x$ posses $p$:
\[ {\rm Her\;}p =_{\rm Df} (\exists y)[y \sqsubset x \land p \in P(y)] \]
A non hereditary property is called an {\em emergent} property.

\item The {\em state} of a thing $x$ is a set of functions from a 
domain of reference $M$ to the set of properties ${\cal P}$.  The set
of the accessible states of a thing $x$ is the {\em lawful state space} of
$x$: $S_{\rm L}(x)$. The state of a thing is represented by a point in
$S_{\rm L}(x)$.

\item A {\em legal statement} is a restriction  upon the state
functions of a given class of things. A {\em natural law}
is a property represented by an empirically corroborated legal statement.

\item The {\em ontological history} $h(x)$ of a thing $x$ is a part of
$S_{\rm L}(x)$ defined by
\[
	h(x) = \{ \langle t, F(t) \rangle | t \in M\} 
\] 
where $t$ is an	element of some auxiliary set $M$,
and $F$ are the functions that represent the properties 
of $x$.

\item There is a single \emph{universal property} of material things
  called \emph{energy} such that the energy of an isolated thing is
  unchanged during its ontological history. 

\item Two things {\em interact} if each of them modifies the history of the
other:
\[
x\Join y\Leftrightarrow h(x\dot{+}y)\neq h(x)\cup h(y)
\]

\item A thing  \(x_{\rm f}\) is a {\em reference frame} for \(x\) iff 
\begin{enumerate} 
  \item $M$ equals the state space of \(x_{\rm f}\), and
  \item  $ h(x  \dot{+} f) = h(x) \cup h(f)$
\end{enumerate}

\item A {\em change} of a thing $x$ is an ordered pair of states:
\[ (s_1, s_2 ) \in E_{\rm L}(x) = S_{\rm L}(x) \times S_{\rm L}(x) \]
A change is called an {\em event}, and the space $E_{\rm L}(x)$ is called
the {\em event space} of $x$.

\item An event $e_1$ {\em precedes} another event $e_2$ if they
compose to give $e_3 \in E_{\rm L}(x)$:
\[
e_1 = ( s_1,s_2 )\; \land\;  e_2 = ( s_2,s_3 )
\Rightarrow  e_3 =  (s_1,s_3 )
\]
\end{enumerate}

The ontology sketched here (due mainly to M. Bunge \cite{Bunge3}. See
also \cite{BungeEnergy}) is realistic, because it assumes the
existence of things endowed with properties, and objective, because it
is free of any reference to cognoscent subjects. 
We will base the axiomatic formulation of the pregeometry of
space-time on this ontology.

\section{Local Time}
\label{sec:Time}

Let us state now the set of hypothesis that introduce the notion of
\emph{local time}. First we assume the existence of an \emph{order
  relation} between states of a given basic thing.%
\footnote{We present an informal classification of axioms in
  mathematical (m),
  ontological (o), semantic (s) and physical (f). This classification
  is non exclusive, i.e. an axiom can belong to more than one class.}
\begin{axiom}[Existence of temporal order (o)]\mbox{}
  For each concrete basic thing $x \in \Theta$ there exist a single
  ordering relation between their states $\leq$.
\end{axiom}

We now give a name to this ordering relation
\begin{axiom}[Denotation of temporal order (s)]\mbox{}
  The set of lawful states of $x$ is \emph{temporally ordered} by the
 $\leq$ relation.
\end{axiom}

The above is a partial order relation: there are pairs of states that
are not ordered by $\leq$; e.g. given an initial condition $(x_0,v_0)$
for a moving particle, there are states  $(x_1,v_1)$ that are not
visited by the particle.
\begin{defin}[Proper history]
  A totally order set of states of $x$ is called a \emph{proper
  history} of $x$.
\end{defin}

The above axioms do not guarantee the existence of a single proper
history: they allow many of them, as in ``The garden of forking
paths'' \cite{BorgesOC}. The following axiom forbids such possibility
\begin{axiom}[Unicity of proper history (o)]
  Each thing has one and only one proper history.
\end{axiom}

\begin{remark}[``Arrow of time'']
  The above axioms describe a kind of ``arrow of time'',  although it is not
  related to irreversibility \cite{sisternaPC}. 
\end{remark}

A proper history is also an ontological history. The parameter $t \in
M$ has not to be continuous. The following axiom, a very strong
version of Heraclitus' hypothesis \emph{Panta rhei}, states that every
thing is changing continuously:
\begin{axiom}[Continuity (o)]
  If the entire set of states of an ontological history is
divided in two subsets $h_{\rm p}$ and $h_{\rm f}$ such that every
state in $h_{\rm p}$ temporally precedes any state in $h_{\rm f}$,
then there exists one and only one state $s_0$ such that $s_1\leq
s_0\leq s_2$, where $s_1\in h_{\rm p}$ and $ s_2\in h_{\rm f}$. 
\end{axiom}

\begin{remark}
  The axiom of continuity is stated in the Dedekind form \cite{RPPCT}.
\end{remark}

\begin{remark}[Continuity in quantum mechanics]
  Although quantum mechanical ``changes of state'' are usually
  considered ``instantaneous'', theory shows that probabilities change
  in a continuous way. The finite width of spectral lines also shows a
  continuous change in time \cite{DanPC}.
\end{remark}

The following theorem can be proved with the standard methods of
analysis \cite{RPPCT}
\begin{theorem}[Real representation]
  Given a \emph{unit change} $(s_0,s_1)$ there exists a bijection
  ${\cal T}: h \leftrightarrow \Re$ such that
  \begin{eqnarray}
    h_1 &=& \{s(\tau)\mid \tau \in \Re\}\\
    s_0 &=& s(0)\\
    s_1 &=& s(1)
  \end{eqnarray}
\end{theorem}

\begin{defin}[Local time]
  The function ${\cal T}$ is called \emph{local time}.
\end{defin}

\begin{remark}
  The unit change $(s_0,s_1)$ is arbitary. It defines an arbitrary ``unit of
  local time'' \cite{sisternaPC}.
\end{remark}

The above theory of local time has an important philosophical
consequence: \emph{becoming}, which is usually conceived as evolution
in time, is here more fundamental than time. The latter is constructed
as an emergent property of a changing (i.e. a  \emph{becoming}) thing.

\section{Simultaneity}
\label{sec:Simultaneity}

In order to introduce the concept of space we shall use the notion of
{\em reflexive action} (or {\em reflex action}) between two
things. Intuitively, a thing $x$ acts on another thing $y$ if the
presence of $x$ disturbs the history of $y$. Events in the real world
seem to happen in such a way that it takes some time for the action of
$x$ to propagate up to $y$. This fact can be used to construct a
relational theory of space {\em \`a la} Leibniz, that is, by taking
space as a set of equitemporal things. It is necessary then to define
the relation of simultaneity between states of things.

\begin{figure}
  \begin{center}
    \includegraphics{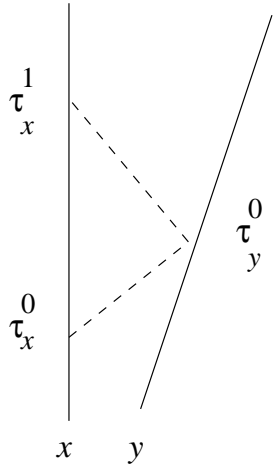}
  \end{center}
  \caption{ Scheme of a reflex action}
  \label{fig:RefAct}
\end{figure}

Let $x$ and $y$ be two things with histories $h(x_{{\tau}} )$ and
$h(y_{{\tau}} )$, respectively, and let us suppose that the action of
$x$ on $y$ starts at ${{\tau}}_x^0$ (See figure \ref{fig:RefAct}). The
history of $y$ will be modified starting from ${{\tau}} _y^0$.  The
proper times are still not related but we can introduce the reflex
action to define the notion of simultaneity. The action of $y$ on $x$,
started at ${{\tau}} _y^0$, will modify $x$ from ${{\tau}} _x^1$
on. The relation ``the action of $x$ on $y$  is reflected
to $x$'' is the reflex action.  Historically, G. Galilei
\cite{GalileoDis} introduced the reflection of a light pulse on a
mirror to measure the speed of light. With this relation we will
define the concept of simultaneity of events that happen on different
basic things (see also \cite{LandauTC}).

Besides we have a second important fact: observation and experiment
suggest that gravitation, whose source is energy, is a universal
interaction, carried by the gravitational field. 

Let us now state the above hypothesis in axiom form. First, we state
\begin{axiom}[Universal interaction]\label{axm:UnivInt}
  Any pair of basic things interact.
\end{axiom}
This extremely strong axiom states not only that there exist no
completely isolated things but that all things are interconnected.

\begin{remark}[Universal inteconection]
  This universal interconnection of things should not be confused with
  ``universal interconnection'' claimed by several mystical
  schools. The present interconnection is possible only through
  physical agents, with no mystical content.
\end{remark}

\begin{remark}[``Accelerated observers'']
  It is possible to model two noninteracting things in Minkowski
  space assuming they are accelerated during an infinite proper
  time. It is easy to see that an infinite energy is necessary to keep
  a constant acceleration, so the model does not represent real
  things, with limited energy supply \cite{DanPC}.
\end{remark}

Now consider the time interval $(\tau_x^1 - \tau_x^0)$. Special
Relativity suggests that it is nonzero, since any action propagates
with a finite speed. We then state
\begin{axiom}[``Finite speed axiom'' (o)]\label{FinSpAx}
  Given two different and separated basic things $x$ and $y$, such as
  in Figure \ref{fig:RefAct}, there exists a minimum positive bound
  for the interval  $(\tau_x^1 - \tau_x^0)$ defined by the reflex
  action. 
\end{axiom}

Now we can define
\begin{defin}[Simultaneity]
  $\tau^0_y$ is simultaneous with  $\tau^{1/2}_x =_{\rm Df} 
  (1/2)(\tau_x^1 + \tau_x^0)$.
\end{defin}

The local times on $x$ and $y$ can be synchronized by the simultaneity
relation. However, as we know from General Relativity, the
simultaneity relation is transitive only in special reference frames
called {\em synchronous} \cite{LandauTC}. We then include the
following axiom:
\begin{axiom}[Synchronizability (f)]
  Given a set of separated basic things $\{x_i\}$ there is an
  assignment of proper times $\tau_i$ such that the relation of
  simultaneity is transitive.
\end{axiom}

With this axiom, the simultaneity relation is an equivalence relation.
Now we can define a first approximation to physical space:
\begin{defin}[Ontic space]
  The equivalence class of states defined by the relation of
  simultaneity on the set of things is the \emph{ontic space} 
  $E_{\rm O}$.
\end{defin}


\section{Universal Time}
\label{sec:Clocks}

The notion of simultaneity allows the analysis of the notion of
\emph{clock}. 
\begin{defin}[Clock]
  A thing $y \in \Theta$ is a \emph{clock} for the thing $x$ if there
  exists an injective function $\psi: S_{\rm L}(y) \to S_{\rm L}(x)$,
  such that $\tau < \tau' \Rightarrow \psi(\tau) < \psi(\tau')$.
\end{defin}
i.e.: the proper time of the clock grows in the same way as the time
of things.

Another much more important concept can be analyzed in the same way
\begin{defin}[Universal time]
  The name \emph{Universal time} applies to the proper time of a
  reference thing that is also a clock.
\end{defin}
From this definition we see that ``universal time'' is frame
dependent in agreement with the results of Special Relativity.

\section{Geometry}
\label{sec:Geometry}

The notion of space we have developed up to now may be called
``Philosopher's space'' \cite{BungeMayne,Bunge3}: there is room for
things there, and separation, but there are no distances. Our next task
is to introduce metric ideas.

\subsection{Pregeometric space}
\label{subsec:PregometricSp}

We shall define distances mimicking the form it is done in relativity
theory: the distance between two simultaneous events is equal to the
time light takes to travel between them multiplied by the velocity of
light. Alas, we do not have space yet, much less electromagnetism or
optics, so we have to take some roundabout. We first introduce $c$
through an axiom:
\begin{axiom}["Light speed" (o)]
  $c$ is a constant (uninterpreted) with suitable dimensions.
\end{axiom}

\begin{remark}
  There is no ambiguity here! The theory of units and dimensions has been
formalized \cite{BungeDim} and the dimensions of distance will depend
on the choice of those of $c$. Only with the development of
electromagnetism it will be possible to interpret $c$ as the speed of
light. This definition is conventionalist, in contrast with the
realistic philosophy adopted \cite{bungePC}.
\end{remark}

Let us recall the definitions of (pseudo)metric and (pseudo)metric
space.
\begin{defin}[Metric (or distance)]
  A \emph{metric} on a set $M$ is a function $d:M\times M \to \Re$
  that satisfies the conditions
  \begin{enumerate}
    \item $d(x,y) = d(y,x)$
    \item $d(x,y) + d(y,z) \ge d(x,z)$
    \item $x=y \Rightarrow d(x,y) = 0$
    \item $d(y,x) = 0  \Rightarrow x=y$
  \end{enumerate}
\end{defin}
If only the first three conditions are satisfied $d(x,y)$ is called a
\emph{pseudo-metric}. 

A set $M$ is a \emph{(pseudo)metric space} if it admits a
(pseudo)metric for every pair of points (elements).

Now consider the reflex action relation (Figure \ref{fig:RefAct}).
We shall first define
\begin{defin}[Ontic distance]
  The \emph{ontic distance} between the simultaneous states at
  $\tau_y^0$ and $\tau_x^{1/2}$ is the function
  \[d(x,y) = \frac{c}{2}\mid  \tau_x^1 - \tau_x^0 \mid\]
\end{defin}

 We still do not know if $d(x,y)$ is a distance (we have given it a
 name, that is all). So we state 
\begin{axiom}[Pseudometric (m)]\label{Pseudometricity}
  The function $d(x,y)$ is a pseudo-metric on the ontic space $E_{\rm
  O}$.
\end{axiom}
With the former axioms one can prove that $E_{\rm O}$ is a
pseudo-metric space and that it can be completed.
\begin{defin}[Pregeometric space]
  \emph{Pregeometric space} $E_{\rm P}$ is the completion of $E_{\rm
  O}$. 
\end{defin}

\begin{remark}
  The function $d(x,y)$ is nonzero for separated things because of
  axiom \ref{FinSpAx}.
\end{remark}

The ontic distance $d(x,y)$ is a pseudo-metric because basic things are
usually bulky and, in the case of gravitational or electromagnetic
fields, they have infinite size.  Axiom \ref{Pseudometricity} only
guarantees that separated things have non-zero distance. 

\subsection{Geometric space}
\label{subsec:GeometricSp}

To build geometric space we have to introduce point-like constructs. 
\begin{defin}[Ontic point]
  Let $\xi \subset \Theta$ be a family of things. We say that $\xi$ is
  a \emph{complete family of united things} if it satisfies:
  \begin{enumerate}
    \item Any two things of $\xi$ are united.
    \item For any thing  $x \not\in \xi$ there is a thing $y \in \xi$
    separated of $x$.
  \end{enumerate}
\end{defin}

\begin{figure}[tb]
  \begin{center}
    \includegraphics{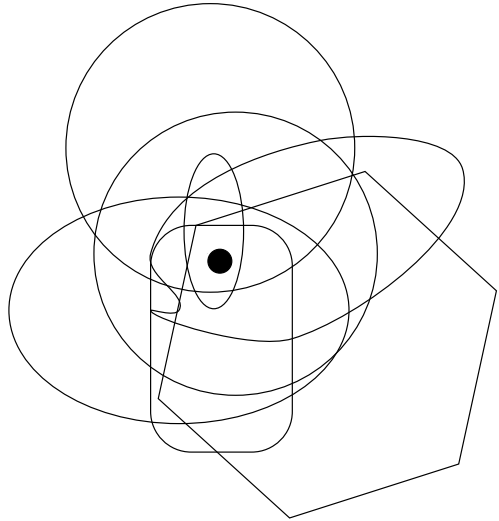}
  \end{center}
  \caption{Scheme of an ontic point: The family of things $\xi$
  ``closes'' around the black dot representing the ontic point.}
  \label{fig:OnPt}
\end{figure}

Now we define a distance between ontic points
\begin{defin}[Distance between ontic points] \mbox{}
  Let $\xi, \eta$ be two ontic points. The \emph{distance between
  ontic points} is 
  \[d_{\rm G}(\xi,\eta) = \sup_{(i,j)}d(x_i,y_j) \]
  where $i \in I, j\in J$ belong to the respective index sets.
\end{defin}

\begin{figure}[tb]
  \begin{center}
    \includegraphics{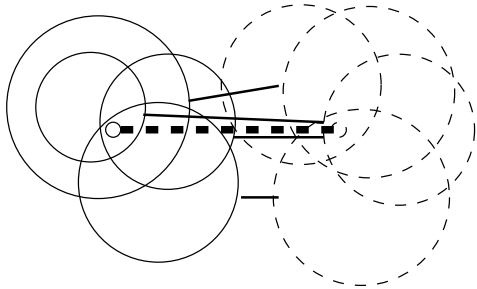}
  \end{center}
  \caption{Construction of the distance between ontic points.}
  \label{fig:SpDist}
\end{figure}

Figure \ref{fig:SpDist} describes in an intuitive way how the pseudo-metric
distances converge to the distance between the two ontic points.

\begin{theorem}[Metricity]
  The set of ontic points is a metric space with distance $d_{\rm
  G}$. 
\end{theorem}
\begin{description}
  \item[Proof:] The first three distance conditions are satisfied
  because $d$ is a pseudo-metric. To show that the fourth is satisfied
  observe that if $\xi \not= \eta$ there are $x_i \wr y_j$ and
  $d(x_i,y_j) > 0$ by axiom \ref{FinSpAx}. So we find
  \begin{eqnarray*}
    \xi \not= \eta &\Rightarrow& d_{\rm G}(\xi,\eta) > 0\\
    d_{\rm G}(\xi,\eta) > 0  &\Rightarrow& \xi = \eta
  \end{eqnarray*}$\Box$
\end{description}

The isometric completion theorem \cite{BourbTR,Thron,Kelley}
guarantees that the metric space of ontic points has a
completion. This justifies the definition
\begin{defin}[Geometric space]
  The completion of ontic space is the \emph{geometric space}
  $E_{\rm G}$.
\end{defin}

\subsection{Euclidean space}
\label{subsec:Euclid}

Finally we need additional hypotheses implying that the structure of
geometric space is euclidean. Blumenthal \cite{Blumenthal} has given a
set of axioms for Euclidean geometry based on the notion of
distance. So we shall assume
\begin{axiom}[Euclidean structure (m)]\mbox{}
    Geometric space satisfies axioms $2, 3, 5', 6'$ and $7$ of
    reference {\rm\cite{Blumenthal}}.
\end{axiom}

An informal description of Blumenthal's axioms is as follows
\begin{description}
  \item[2 and 3:] There are at least three aligned points.
  \item[5':] There are three unaligned points.
  \item[6':] There are four non coplanar points.
  \item[7:] There is no fourth dimension.
\end{description}
The exact formulation of these axioms uses only the notion of
distance. 

On the other hand, the fourth axiom of Blumenthal is a theorem in this
formulation:
\begin{theorem}[Completeness]\label{the:Completeness}
  Geometric space $E_{\rm G}$ is complete.
\end{theorem}
which follows from the isometric completion theorem.  From the
above, the following result can be derived \cite{Blumenthal}:
\begin{theorem}[Euclidicity]
  Geometric space $E_{\rm G}$ is euclidean.
\end{theorem}


 Theorem \ref{the:Completeness} has a deep ontological
 consequence. Since ontic space $E_{\rm O}$ is dense in geometrical
 space $E_{\rm G}$ we derive the
 \begin{theorem}[Aristotle-Leibniz]
   Ontic space is a \emph{plenum}.
 \end{theorem}
that is: there are concrete things everywhere. 

\begin{remark}
  \label{rem:Plenum}
  This theorem is, in
spite of appearances, in agreement with modern physics.  Indeed, the
\emph{plenum} hypothesis (introduced by Aristotle and later supported
by Leibniz) is confirmed in Quantum Physics, and it leads to the
prediction of a plethora of vacuum phenomena (like the Casimir
effect), in good agreement with observation.
\end{remark}

\begin{remark}
  Let us remark that we have not assumed the existence of a
\emph{plenum}: is a consequence that the ontic space is dense in
$E_{\rm G}$. Neither we have assumed that ontic space $E_{\rm O}$ is
euclidean, but that it is dense in an Euclidean space.
\end{remark}

\begin{remark}
  Remark \ref{rem:Plenum} suggests that quantum mechanics is a
  necessary extension of classical mechanics to get a
  \emph{plenum}. This is not true, since it is possible to ``fill''
  the space with fluids, such as \emph{dark matter} or \emph{``dark
    energy''}, as it is assumed in modern cosmology \cite{sisternaPC}.
\end{remark}

Our final axiom is a semantic one, stating the interpretation of the
geometric space
\begin{axiom}[Physical space (s)]
  $E_{\rm G}$ represents physical space $E_{\rm Ph}$.
\end{axiom}

This axiom closes the present theory of space-time.

\section{Conclusion}
\label{sec:Concl}

	In the present theory, space-time is not a thing but a
substantial property of the largest system of things, the world
\(\Box\), emerging from the set of the relational properties of basic
things. Thus, any existential quantification over space-time can be
translated into quantification over basic things. This shows that
space-time has no ontological independence, but it is the product of
the interrelation between basic ontological building blocks. For
instance, rather than stating ``space-time possesses a metric'', it
should be said: ``the evolution of interacting things can be described
attributing a metric tensor to their spatiotemporal
relationships''. In the present theory, however, space-time is
interpreted in an strictly materialistic and Leibnitzian sense: it is
an order of successive material coexistents.


We have mentioned above some simple philosophical consequences of this
theory: becoming is more fundamental than time, and space (space-time,
indeed) is a \emph{plenum}. 

We have exposed a materialistic relational theory of space-time, that
carries out the program initiated by Leibniz, and provides a
protophysical basis consistent with any rigorous formulation of
General Relativity. Space-time is constructed from general concepts
which are common to any consistent scientific theory. The particular
hypothesis used for the construction have been taken from well
corroborated scientific facts. It is shown, consequently, that there
is no need for positing the independent existence of space-time over
the set of individual things.

\section*{Acknowledgments}

I am indebted to M. Bunge, G. Romero, P. Sisterna and  D. Sudarsky for
valuable criticism, comment and advice.

\bibliography{filosof,gener,PrivComm}

\end{document}